\documentclass[notitlepage, aps, prl, reprint, superscriptaddress, amsmath, amssymb]{revtex4-2}
\pdfoutput=1
\usepackage{amsmath}          
\usepackage{amsfonts}          
\usepackage{mathtools}        
\usepackage{bbm}
\usepackage{float}            
\usepackage{graphicx}         
\usepackage[dvipsnames]{xcolor}
\usepackage{cancel}           
\usepackage{relsize}          
\usepackage{epstopdf}
\usepackage{soul}
\usepackage{hyperref}
\hypersetup{
	linkcolor=blue,
	citecolor=blue,
	filecolor=blue,
	urlcolor=blue,
	colorlinks=true
}

\usepackage{pifont}      
\usepackage{color}       
\usepackage{framed}		 
\definecolor{shadecolor}{gray}{0.965}
\usepackage{siunitx}	 

\renewcommand{\v}[1]{\ensuremath{\boldsymbol{\mathbf{#1}}}} 

\newcommand{\UIUCPHYS}[0]{Department of Physics, University of Illinois at Urbana-Champaign, Urbana, IL 61801, USA}
\newcommand{\UIUCECE}[0]{Department of Electrical and Computer Engineering, University of Illinois at Urbana-Champaign, Urbana, IL 61801, USA}
\newcommand{\UIUCMNTL}[0]{Micro and Nanotechnology Laboratory, University of Illinois at Urbana-Champaign, 208 N. Wright Street, Urbana IL 61801, USA}
\newcommand{\STANFORDECE}[0]{Department of Electrical Engineering, Stanford University, Stanford, California 94305, USA}
\newcommand{\IBM}[0]{IBM Research at Albany Nanotech, 257 Fuller Road, Albany, New York 12203, USA}

\allowdisplaybreaks 

\begin{document}

\title{Topology and Observables of the Non-Hermitian Chern Insulator}

\author{Mark R. Hirsbrunner}\email{hrsbrnn2@illinois.edu}\affiliation{\UIUCPHYS}\affiliation{\UIUCMNTL}
\author{Timothy M. Philip}\altaffiliation[Current address: ]{\IBM}\affiliation{\UIUCMNTL}\affiliation{\UIUCECE}
\author{Matthew J. Gilbert}\affiliation{\UIUCMNTL}\affiliation{\UIUCECE}\affiliation{\STANFORDECE}

\begin{abstract}
Topology plays a central role in nearly all disciplines of physics, yet its applications have so far been restricted to closed, lossless systems in thermodynamic equilibrium. Given that many physical systems are open and may include gain and loss mechanisms, there is an eminent need to reexamine topology within the context of non-Hermitian theories that describe open, lossy systems. The recent generalization of the Chern number to non-Hermitian Hamiltonians initiated this reexamination; however, there is so far no established connection between a non-Hermitian topological invariant and the quantization of an observable. In this work, we show that no such relationship exists between the Chern number of non-Hermitian bands and the quantization of the Hall conductivity. Using field theoretical techniques, we derive an exact expression for the non-quantized Hall conductivity of a generic two-level non-Hermitian Hamiltonian. Furthermore, we calculate the longitudinal and Hall conductivities of a non-Hermitian Hamiltonian with a finite Chern number to explicitly demonstrate the disconnect between the Hall conductivity and the Chern number. These results demonstrate that the Chern number does not provide a \emph{physically meaningful} classification of non-Hermitian Hamiltonians.
\end{abstract}

\maketitle


The topological classification of matter represents a significant enhancement in our understanding of the physical properties of a great variety of systems, both classical~\cite{susstrunk_observation_2015,huber_topological_2016,susstrunk_classification_2016} and quantum-mechanical in nature~\cite{ryu_topological_2010,kitaev_periodic_2009}. Of central importance within the topological classification of matter is the identification of topological invariants, which are quantities that remain unchanged in the presence of symmetry-allowed perturbations~\cite{hasan_colloquium_2010,ryu_classification_2008,bernevig_topological_2013,Qi-RMP-2011}. While the topological classification of matter has enjoyed much success, its achievements have to date been limited to idealized closed systems, as described by conventional Hermitian Hamiltonians. Nonetheless, most physical systems are more aptly described as open, defined by a connection to large reservoirs of additional states. Proper theoretical descriptions of open systems must include mechanisms of both loss and gain that account for the flow of energy and particles between the system and additional reservoirs~\cite{rotter_review_2015,rotter_non-hermitian_2009,eleuch_open_2015}. The inclusion of gain and loss mechanisms necessitates a non-Hermitian Hamiltonian, whose complex eigenvalues induce finite quasiparticle lifetimes. Non-Hermitian Hamiltonians permit many topological phenomena that are discordant with their Hermitian counterparts including: exceptional points, lines, and surfaces at which two eigenvectors merge into one~\cite{shen_topological_2018,san-jose_majorana_2016,martinez_alvarez_non-hermitian_2018,moors_disorder-driven_2019,budich_symmetry-protected_2019,okugawa_topological_2019,yang_non-hermitian_2019}, unidirectional optical transport~\cite{longhi_robust_2015,feng_experimental_2013}, bulk Fermi arcs~\cite{zhou_observation_2018}, expanded topological classifications~\cite{lieu_topological_2018,lieu_topological_2018-1,gong_topological_2018,kawabata_topological_2019}, and a modified bulk-boundary correspondence~\cite{lee_anomalous_2016,yao_non-hermitian_2018,kunst_biorthogonal_2018,kawabata_anomalous_2018,lang_effects_2018,martinez_alvarez_non-hermitian_2018,xiong_why_2018,yao_non-hermitian_2018,yao_edge_2018,jin_bulk-boundary_2019,lee_anatomy_2018,herviou_restoring_2018}.

The breakdown of the conventional bulk-boundary correspondence in non-Hermitian topological Hamiltonians calls for the reexamination of other predictions of topology in non-Hermitian systems. One of the sacrosanct tenants of topological physics is the connection between topological invariants and quantized observables. Within the context of gapped Hermitian Hamiltonians, the Chern number of the energy bands is equivalent to the number of chiral edge states, as required by the bulk-boundary correspondence~\cite{hasan_colloquium_2010, bernevig_topological_2013,Qi-RMP-2011}. The connection between the number of edge states and the Chern number, in turn, leads to a quantized Hall conductivity in units of $e^2/h$~\cite{thouless_quantized_1982}. The Chern number thus provides both a mathematical classification of the Hamiltonian and a physical characterization of the resultant phase.

In this work, we demonstrate that the intimate link between the Hall conductivity and the Chern number no longer holds in a non-Hermitian Chern insulator. Specifically, we show that the Chern-Simons response coefficient in the the effective action of a non-Hermitian Chern insulator is not quantized despite the quantization of the Chern number.  We derive an exact expression for the non-quantized Hall conductivity of a generic non-Hermitian two level system, in which the anti-Hermitian component may have arbitrary dependence on momentum. As a concrete demonstration of the disconnect between topology and observable, we calculate the Hall conductivity of a non-Hermitian massive Dirac Hamiltonian in $(2+1)$-D, which has a non-zero Chern number. We find that the Hall conductivity may be continuously tuned from the quantized value $\sigma_{xy} = e^2/2h$ to $\sigma_{xy}=0$ by increasing the quasiparticle broadening, or equivalently reducing the quasiparticle lifetime, without changing the Chern number, proving that the Hermitian and non-Hermitian Chern insulators are not adiabatically connected.

We begin by examining the Hall conductivity of a gapped, translationally invariant Hermitian system in $(2+1)$-D. As calculated via the Kubo formula, the Hall conductivity is
\begin{equation}
\sigma_{xy} = \frac{ie^2\hbar}{V}\sum_{m,n}\left(f_m-f_n\right)\frac{\left\langle m\right|\hat{v}_x\left|n\right\rangle\left\langle n\right|\hat{v}_y\left|m\right\rangle}{\left(\epsilon_m-\epsilon_n\right)^2},
\label{eq:HermKubo}
\end{equation}
where $V$ is the volume of the system, $f_i=f(\epsilon_i)$ is the Fermi-Dirac distribution function, $\hat{v}_i = \frac{1}{\hbar}d\hat{H}/dk_i$ are the velocity operators, $\epsilon_n$ and $\left|n\right\rangle$ are the energies and eigenstates of the Hamiltonian $\hat{H}$, and $m$, $n$ index the eigenstates of $\hat{H}$. For a gapped Hamiltonian, Eq.~\eqref{eq:HermKubo} may be recast as an integral of the Berry curvature of the occupied bands over the Brillouin zone~\cite{thouless_quantized_1982},
\begin{equation}
\sigma_{xy} = \frac{ie^2}{2\pi h}\sum_{q \in occ}\int_{\text{BZ}}\epsilon_{ij}\left\langle\partial_i \Psi_{q}(\mathbf{k})\right|\left.\partial_j \Psi_{q}(\mathbf{k})\right\rangle d^2\mathbf{k} ,
\label{eq:hallbc}
\end{equation}
where $q$ indexes the occupied bands. The integral of the Berry curvature, $\epsilon_{ij}\left\langle\partial_i \Psi_{n}(\mathbf{k})\right|\left.\partial_j \Psi_{n}(\mathbf{k})\right\rangle$, over the Brillouin zone defines the Chern number. The Hall conductivity is proportional to the Chern number and is quantized to \smash{$\sigma_{xy}=ne^{2}/h$}, where $n\in \mathbb{Z}$.

However, Eqs.~\eqref{eq:HermKubo} and~\eqref{eq:hallbc} fail for non-Hermitian Hamiltonians as they explicitly rely upon the ability to distinguish occupied and unoccupied eigenstates. The failure is caused by the complex energy eigenvalues possessed by non-Hermitian Hamiltonians, for which the Fermi distribution does not produce occupation probabilities. Although the Kubo approach to the Hall conductivity is inappropriate for non-Hermitian Hamiltonians~\footnote{The Hall conductivity of non-Hermitian systems has previously been studied via a modification of linear response theory~\cite{shen_hall_2014}}, we may still construct the effective action of an external $U(1)$ gauge field to obtain the Hall conductivity, an approach that is valid for both free and interacting theories~\cite{qi_topological_2008,wang_topological_2010}. The Hall response of a gapped system is contained in the topological Chern-Simons term of the effective action~\footnote{If there is a finite spectral density in the gap, as is the case for non-Hermitian Hamiltonians, other higher-order and non-local terms are allowed and can contribute a non-quantized Hall conductivity. We do not address such terms in this work.},
 \begin{equation}
S_{\text{CS}}[A] = \frac{C_{\text{CS}}}{4\pi}\int\,d^3x\,\epsilon^{\mu\nu\rho}A_\mu\partial_\nu A_\rho,
\end{equation}
where $A_\mu$ is the electromagnetic vector potential. The Hall conductivity is proportional to the response coefficient, $\sigma_{xy} = C_{\text{CS}}e^2/h$, which is calculated from the linear, antisymmetric part of the polarization tensor as~\cite{ishikawa_magnetic_1986,ishikawa_microscopic_1987}
\begin{equation}
\sigma_{xy} = \frac{e^2}{h}\frac{\epsilon^{\mu\nu\rho}}{24\pi^2}\int\, d^3p\,\text{Tr}\left[G\frac{\partial G^{-1}}{\partial p_\mu} G\frac{\partial G^{-1}}{\partial p_\nu} G\frac{\partial G^{-1}}{\partial p_\rho}\right], 
\label{eq:IMdef}
\end{equation}
where $p=(\omega,k_x,k_y)$, the frequency $\omega$ is integrated along the imaginary axis of the complex plane, and $G$ is the Matsubara Green function. The Matsubara Green function in Eq.~\eqref{eq:IMdef} is defined as
\begin{equation}
G(\omega,\v{k}) = \left[\omega-H(\v{k})-\Sigma(\omega,\v{k})\right]^{-1},
\end{equation}
where $H$ is the Hamiltonian and $\Sigma$ is the self-energy. The self-energy accounts for the presence of energy exchange between the system and reservoirs as well as dissipative interactions, both of which combine to imbue the quasiparticles with a finite lifetime. The electromagnetic response of the Chern-Simons term is identical to the Kubo formula for the Hall conductivity and thus results in an identical expression for the quantized Hall conductivity $\sigma_{xy}=C_{\text{CS}}e^2/h$.

To clearly understand the topological quantization of Eq.~\eqref{eq:IMdef}, we recognize that the Green function represents a homeomorphism, a continuous bijection with a continuous inverse, between $(2+1)$-D momentum space and the general linear group $GL(N,\mathbb{C})$, where $N$ is the number of energy bands. Let us first consider the continuum case, in which momentum space is isomorphic to $\mathbb{R}^3$. Since the Green function approaches zero in the limits $k\rightarrow\infty$ and $\omega\rightarrow\infty$, we can compactify momentum space into the three-sphere $S^3$ by adding a point at infinity. With the point at infinity, the Green function now defines a three-loop in $GL(N,\mathbb{C})$~\cite{nakahara_geometry_2003}. Therefore, the Green function is an element of the third homotopy group of the general linear group, $\pi_3\left(GL(N,\mathbb{C})\right)$, which is isomorphic to $\mathbb{Z}$. In the lattice case, momentum space can be compactified into a pinched torus, whose third homotopy group is also isomorphic to $\mathbb{Z}$~\cite{ishikawa_magnetic_1986}. Eq.~\eqref{eq:IMdef} identifies to which element of $\mathbb{Z}$ the Green function corresponds, guaranteeing the integer quantization of the Hall conductivity in the Chern insulator. 


In order to evaluate Eq.~\eqref{eq:IMdef}, we must construct the requisite non-Hermitian Green function. Consider a general non-Hermitian Hamiltonian, written as
\begin{equation}
H(\mathbf{k}) = H_0(\mathbf{k}) + \Gamma(\mathbf{k}),
\end{equation}
where the Hamiltonian has been broken up into Hermitian, $H_0=H_0^\dagger$, and anti-Hermitian, $\Gamma=-\Gamma^\dagger$, components. In this formulation, the anti-Hermitian component is relegated to a self-energy term, giving the Matsubara Green function
\begin{equation}
G(\omega,\mathbf{k}) = \frac{1}{\omega - H_0(\mathbf{k}) -\Gamma(\mathbf{k})\text{sgn}(\text{Im }\omega)},
\end{equation}
where $\Sigma(\omega,\mathbf{k}) = \Gamma(\mathbf{k})\text{sgn}(\text{Im }\omega)$. In order to preserve causality,
we require the eigenvalues of $\Gamma(\mathbf{k})$ to lie on the negative imaginary axis~\cite{supplement}.

\begin{figure}
\centering
\includegraphics[width=0.5\textwidth]{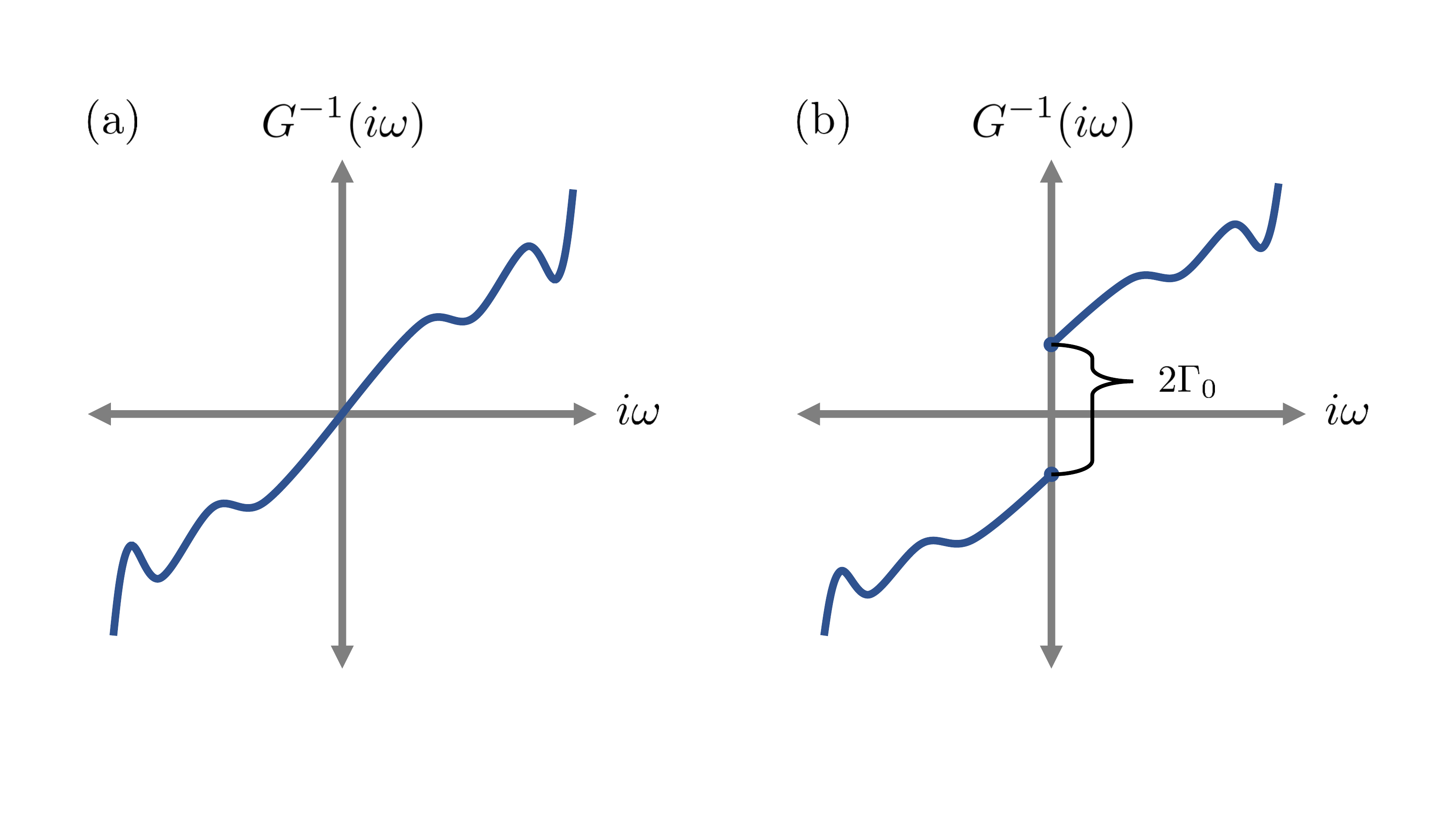}
\caption{Schematic representation of the inverse of a (a) conventional Green function and (b) non-Hermitian Green function, as a function of $i\omega$. The non-Hermitian self-energy $\Gamma(\mathbf{k})\text{sgn}(\text{Im }\omega)$ causes a discontinuity of magnitude $2\Gamma(\mathbf{k})=2\Gamma_0$ at $\omega=0$.}
\label{fig:Discont}
\end{figure}

The salient feature of the non-Hermitian Green function is the frequency dependence of the self-energy. The self-energy depends on $\omega$ only via the signum function because it has been extracted from the Hamiltonian, which has no dependence on $\omega$. The frequency dependence of the self-energy induces a discontinuity in every non-Hermitian Green function at $\omega=0$, as demonstrated in the schematic in Fig.~\ref{fig:Discont}. This discontinuity is avoided by the self-energy of most common interactions by an additional dependence on $\omega$ that sets the magnitude of the self-energy to zero at $\omega=0$. Such a discontinuous Green function is not a homeomorphism and cannot be identified via Eq.~\eqref{eq:IMdef} with an element of $\pi_{3}(GL(N,\mathbb{C}))\cong\mathbb{Z}$. As a result, the Hall conductivity of non-Hermitian Hamiltonians is neither a topological invariant nor quantized.

The topological invariance of Eq.~\eqref{eq:IMdef} may be proven by demonstrating that the variation in the Hall conductivity induced by a variation of the Green function is identically zero. Under the general distortion $G\rightarrow G+\delta G$, the variation in the Hall conductivity is written as~\cite{supplement}
\begin{equation}
\delta \sigma_{xy} = -\frac{e^2}{h}\frac{\epsilon^{\mu\nu\rho}}{24\pi^2}\int\, d^3p\,\partial_\mu \text{Tr}\left[\delta G\partial_\nu G^{-1} G\partial_\rho G^{-1}\right]. \label{eq:zero_variation1}
\end{equation}
For a smooth, continuous Green function, this expression can be recast as a surface integral via the divergence theorem. Since the distortion $\delta G$ must go to zero at the boundary $(\omega\rightarrow\pm\infty)$, the variation is identically zero and the Hall conductivity is a topological invariant. However, the divergence theorem only applies to continuous functions, and thus cannot be used to evaluate the variation of non-Hermitian Green functions. Since $\delta G$ is arbitrary, the integral can effectively take any value, thus the variation is finite and the Hall conductivity is not a topological invariant. The above discussion is completely general to any non-Hermitian Hamiltonian as we have not \emph{a priori} assumed any particular form of the self-energy. 

To illustrate the impact of a discontinuity in the Green function, we consider a general diagonal self-energy $\Sigma(\omega,\mathbf{k}) = -i\Gamma_0(\omega,\mathbf{k})\text{sgn}(\text{Im }\omega)I$, where $\Gamma_0(\omega,\mathbf{k})$ is positive and real. This self-energy can be substituted into the frequency variable in Eq.~\eqref{eq:zero_variation1}, resulting in a variation in the Hall conductivity of the form~\cite{supplement}
\begin{equation}
\begin{aligned}
\delta\sigma_{xy} =& \frac{e^2}{h}\frac{\epsilon^{ij}}{24\pi^2} \times \\
&\left.\int d^2k\, \text{Tr}\left[\delta G\partial_i G_0^{-1}G_0\partial_j G_0^{-1}\right]\right|_{\omega'=-i\Gamma_0(0,\mathbf{k})}^{\omega'=i\Gamma_0(0,\mathbf{k})}\label{eq:finite_variation}
\end{aligned},
\end{equation}
where $G_0$ is the bare Green function with no self-energy and the indices $i$ and $j$ span the momenta $k_x$ and $k_y$. If $\Gamma_0(0,\mathbf{k})=0$, this expression is zero and the Hall conductivity is a topological invariant. The self-energy arising from any Fermi-liquid interaction, for example, is identically zero at $\omega=0$, and leaves the Hall conductivity an invariant. However, since $\Sigma(\omega,\mathbf{k})=\Gamma(\mathbf{k})$ has no frequency dependence for non-Hermitian Green functions, the terms in this expression do not cancel each other and the result is finite. Since $\delta G$ is arbitrary, it is not restricted to be zero at $\omega=\pm i\Gamma_0(\omega,\mathbf{k})$, thus the variation in the Hall conductivity is no longer guaranteed to be zero and the Hall conductivity is not a topological invariant.

We can further understand the non-quantization of the Hall conductivity of the non-Hermitian Chern insulator by considering a generic, gapped, two level system described by the Hamiltonian
\begin{equation}
H(\mathbf{k}) = d_0(\v{k})\sigma_0 + \v{d}(\v{k})\cdot\v{\sigma},
\end{equation}
where $d_0,d_i\in\mathbb{R}$ and $\v{\sigma}$ is a vector of the Pauli matrices. The topological quantization of the Hall conductivity is made clear by expressing it as~\cite{fruchart_introduction_2013}
\begin{equation}
\sigma_{xy} = \frac{e^2}{h}\int\frac{d^2k}{4\pi}\hat{d}\cdot\left(\frac{\partial \hat{d}}{\partial k_x}\times\frac{\partial \hat{d}}{\partial k_y}\right). \label{eq:2levelChern}
\end{equation}
The integral in this expression measures the solid-angle that the vector $\v{d}(\mathbf{k})$ sweeps out on $S^2$ as the momentum is integrated over the Brillouin zone. This geometric quantity must be an integer, and is formally equivalent to the Chern number.

The non-Hermitian generalization of this Hamiltonian is
\begin{equation}
	H(\mathbf{k}) = \left(d_0(\v{k}) + i\Gamma_0(\v{k})\right)\sigma_0+\left(\v{d}(\v{k}) + i\v{\Gamma}(\v{k})\right)\cdot\v{\sigma},
\end{equation}
where $\v{\Gamma}(\v{k})$ is a vector of real numbers and must satisfy the requirement that the eigenvalues of $H(\v{k})$ have negative imaginary components. In order to make the following calculation more transparent, we suppress any momentum dependence and use the following definitions: $b_0 = d_0 + i\Gamma_0$, $\v{b} = \v{d}+i\v{\Gamma}$, and $b = \sqrt{\v{b}\cdot\v{b}}$. Using Eq.~\eqref{eq:IMdef}, we find the Hall conductivity of a generic two-level non-Hermitian Hamiltonian to be~\cite{supplement}
\begin{equation}
\begin{aligned}
&\sigma_{xy} = -\frac{e^2}{h}\int \frac{d^2k}{2\pi^2}\text{Re}\left[\hat{b}\cdot\left(\frac{\partial \hat{b}}{\partial k_x}\times\frac{\partial \hat{b}}{\partial k_y}\right) \times \right. \\
&\quad\left.\left(\frac{\pi}{2}\text{sgn}(\text{Re }b)  - \frac{ibb_0}{b^2-b_0^2} - i\text{ arctanh}\left(\frac{b_0}{b}\right)\right)\right] \label{eq:2level_Hall}.
\end{aligned}
\end{equation}
The infinitesimal angle swept out by the vector $\v{b}(\mathbf{k})$ is now multiplied by a function of the momentum, thus the integral does not count the number of times $\v{b}(k)$ covers the sphere. This compact expression for the Hall conductivity as an integral over the Brillouin zone makes manifest the absence of a topological interpretation. 

\begin{figure}
\centering
\includegraphics[width=0.5\textwidth]{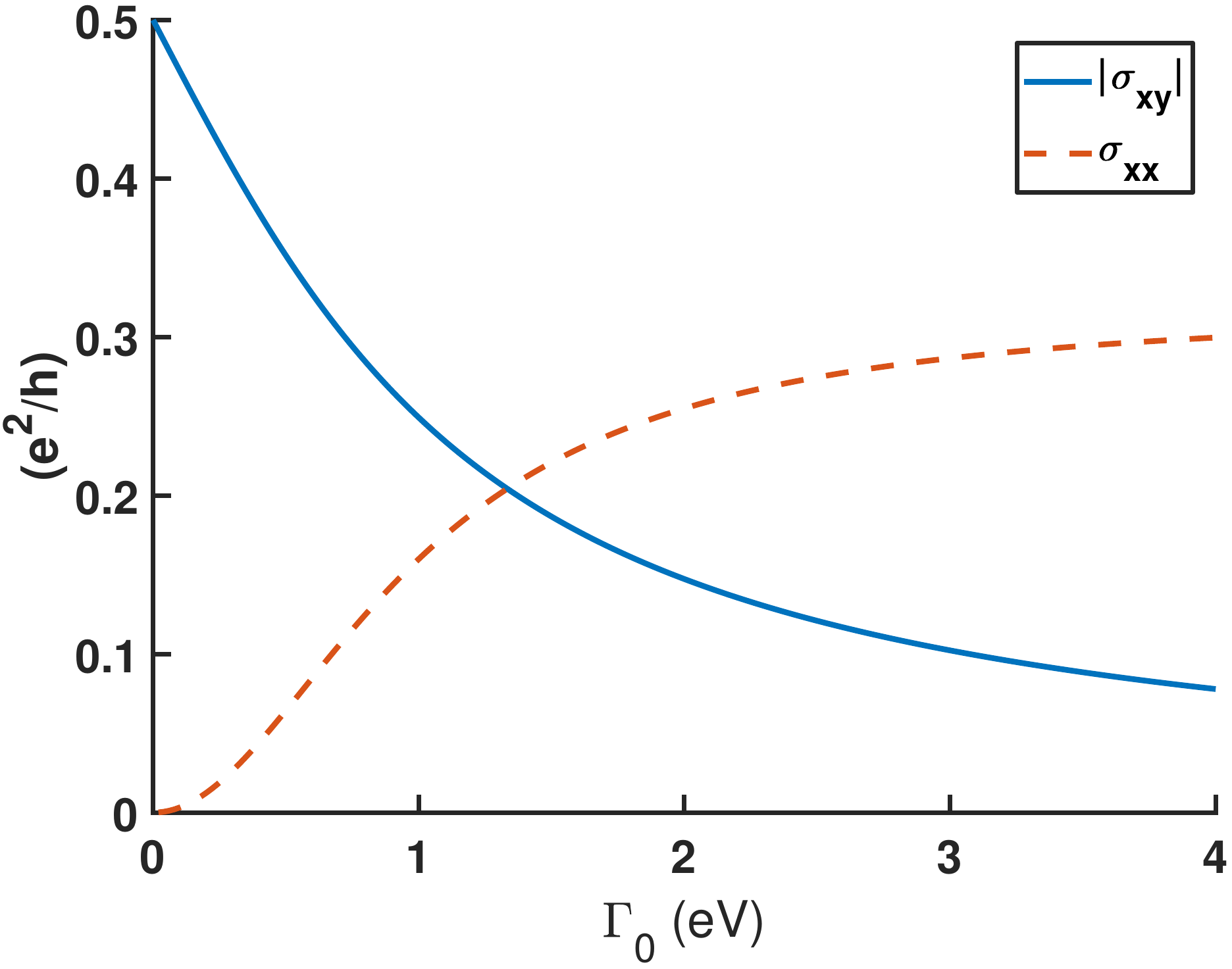}
\caption{The longitudinal conductivity and magnitude of the Hall conductivity for the non-Hermitian Chern insulator as a function of the broadening, $\Gamma_0$, with $\mu=0.1$ eV and $M=1$ eV. The Hall conductivity monotonically decreases from $\left|\sigma_{xy}\right|=e^2/2h$ to $\sigma_{xy} = 0$, while the longitudinal conductivity monotonically increases from $\sigma_{xx}=0$ to $\sigma_{xx}=\frac{1}{\pi} e^2/h$.}
\label{fig:Hall_Cond}
\end{figure}

To further elucidate the disconnection between Chern number and bulk topological invariant, we now analyze the Hall conductivity of a model non-Hermitian Chern insulator in detail. To this end, we utilize an inversion-symmetric massive Dirac Hamiltonian, given by
\begin{equation}
H_0(\v{k}) = -\mu\sigma_0 + \nu_F\v{k}\cdot\v{\sigma} + M\sigma_z,
\end{equation}
where $\mu$ is the chemical potential, $\nu_F$ is the Fermi velocity, and $M>|\mu|$ is the energy gap. When the chemical potential is within the energy gap, the massive Dirac Hamiltonian has a vanishing longitudinal conductance and a Chern number $C=-\frac{1}{2}$~\footnote{The half-integer quantized Hall conductivity is an exception to the usual integer quantization resulting from the meron configuration of the spin texture for a massive Dirac Hamiltonian.}, corresponding to a half-quantized Hall conductance $\sigma_{xy} = -e^2/2h$~\cite{ludwig_integer_1994}. We generalize this model to a non-Hermitian Chern insulator by adding a constant diagonal imaginary term that respects the same symmetry as the Hamiltonian,
\begin{equation}
H(\v{k}) = -(\mu + i\Gamma_0)\sigma_0 + \nu_F\v{k}\cdot\v{\sigma} + M\sigma_z.
\end{equation}
As the anti-Hermitian component of the Hamiltonian, $\Gamma(\mathbf{k})=-i\Gamma_0\sigma_0$, is proportional to the identity matrix, the eigenvectors of the Hamiltonian and the Chern number are unchanged from the Hermitian case. Using Eq.~\eqref{eq:IMdef}, we calculate the Hall conductivity of this non-Hermitian massive Dirac Hamiltonian to be~\cite{supplement}
\begin{equation}
\sigma_{xy} = \frac{e^2}{h}\frac{M}{2\pi|M|}\left[\arctan\left(\frac{\mu^2+\Gamma_0^2-M^2}{2\Gamma_0|M|}\right) - \frac{\pi}{2}\right],
\label{eq:TI_Hall}
\end{equation}
in agreement with previous results on non-Hermitian massive Dirac systems~\cite{Note4}. Eq.~\eqref{eq:TI_Hall} yields the properly quantized value $\sigma_{xy} = -e^2/2h$ in the Hermitian limit $\Gamma_0\rightarrow0$, as expected. However, for any finite value of broadening, $\Gamma_0$, the Hall conductivity is reduced from its Hermitian value, as shown in Fig.~\ref{fig:Hall_Cond}, approaching $\sigma_{xy} = 0$ as $\Gamma_0\rightarrow\infty$. Consequently, we see that the Hall conductivity of the non-Hermitian Chern insulator may be continuously varied by manipulating the broadening without changing the Chern number.

\footnotetext[4]{There are other contributions to the Hall conductivity of this system, such as the Fermi surface term~\cite{philip_loss_2018}.}

With the loss of quantization in the Hall conductivity, one expects an associated response in the longitudinal conductivity~\cite{Klitzing-RMP-1986}. As the broadening increases, a finite spectral density develops in the gap, allowing for conduction through the bulk of the system. We may write the longitudinal conductivity in terms of Green functions as~\cite{ryu_landauer_2007}
\begin{equation}
\begin{aligned}
\sigma_{xx} = -\frac{e^2}{2h}\int\frac{d^2k}{(2\pi)^2}\text{Tr}&\left[\text{Im }G^A(\mathbf{k},0)\frac{\partial H(\mathbf{k})}{\partial {k_x}}\right. \\
\times&\left.\text{Im }G^A(-\mathbf{k},0)\frac{\partial H(\mathbf{k})}{\partial {k_x}}\right],
\label{eq:longcondef}
\end{aligned}
\end{equation}
where $G^A(\mathbf{k},\omega)$ is the advanced Green function~\cite{supplement}. Substituting the Green function of this non-Hermitian Chern insulator into Eq.~\eqref{eq:longcondef} gives the conductivity
\begin{equation}
\begin{aligned}
\sigma_{xx} = &\frac{\mu^2+\Gamma_0^2-M^2}{4\pi\Gamma_0\mu}\left[\frac{2\Gamma_0\mu}{\mu^2+\Gamma_0^2-M^2}\right. \\
&\quad\quad+
\left.\text{arctan}\left(\frac{2\Gamma_0\mu}{M^2+\Gamma_0^2-\mu^2}\right)\right]\frac{e^2}{h},
\end{aligned}
\label{eq:longcon2}
\end{equation}
In examining Eq.~\eqref{eq:longcon2}, we observe that in the Hermitian limit, $\Gamma_0\rightarrow0$, the longitudinal conductivity goes to zero, as it must for a Hermitian gapped system. In both the massless limit, $M\rightarrow0$, and in the limit of infinite broadening, $\Gamma_0\rightarrow\infty$, the conductivity approaches the theoretical minimum conductivity of a single Dirac cone~\cite{ryu_landauer_2007},
\begin{equation}
\lim_{M\rightarrow0}\sigma_{xx}=\lim_{\Gamma_0\rightarrow\infty}\sigma_{xx}=\frac{e^2}{\pi h}.
\end{equation}
Between these two limits, the longitudinal conductivity remains finite, indicating that the loss of the quantization in the Hall conductivity is indeed associated with a finite longitudinal conductivity.

A natural extension is to consider non-Hermitian systems in dimensions higher than $(2+1)$. To this point, we consider the $(4+1)$-D quantum Hall insulator, a higher-dimensional analogue of the Chern insulator that is described by the Chern-Simons action
\begin{equation}
	S_{\text{eff}} = \frac{C_2}{24\pi^2}\int d^4xdt\epsilon^{\mu\nu\rho\sigma\tau}A_\mu\partial_\nu A_\rho\partial_\sigma A_\tau,
\end{equation}
which corresponds a non-linear Hall response of the form~\cite{qi_topological_2008,zhang_four-dimensional_2001}
\begin{equation}
		j^\mu = \frac{C_2}{8\pi^2}\epsilon^{\mu\nu\rho\sigma\tau}\partial_\nu A_\rho\partial_\sigma A_\tau.
\end{equation}
Here the coefficient $C_2$ is the second Chern number of the non-Abelian Berry phase~\cite{qi_topological_2008}, which may be expressed via Green functions as
\begin{equation}
\begin{aligned}
	C_2 = -\frac{\pi^2}{15}\epsilon^{\mu\nu\rho\sigma\tau}\int\, &\frac{d^5p}{(2\pi)^5}\,\text{Tr}\left[G\frac{\partial G^{-1}}{\partial p_\mu} G\frac{\partial G^{-1}}{\partial p_\nu} \right. \\
	&\quad\left.G\frac{\partial G^{-1}}{\partial p_\rho} G\frac{\partial G^{-1}}{\partial p_\sigma} G\frac{\partial G^{-1}}{\partial p_\tau}\right].
\end{aligned}
\end{equation}
This integral is a higher-dimensional form of the topological invariant in Eq.~\eqref{eq:IMdef}, as it identifies the Green function with an element of $\pi_5(GL(N,\mathbb{C}))=\mathbb{Z}$, resulting in a quantized non-linear Hall response. The discontinuity in non-Hermitian Green functions invalidates this topological quantization argument, as it did in the $(2+1)$-D case, again leading to a disconnect between a topological invariant and a quantized observable in a higher-dimensional Chern insulator.

The fact that non-Hermiticity results in a non-quantized Hall conductivity despite a quantized Chern number seems to be directly at odds with the clear experimental observations of the quantized Hall conductivity in magnetically-doped three-dimensional time-reversal invariant topological insulators~\cite{Chang2015a,Chang2013,Kou2014,Bestwick2015,Fox2018}. Such a mesoscopic system is generally open and disordered, meaning it may be best described by a non-Hermitian Hamiltonian that accounts for finite lifetimes. The reason that the disconnect between topological observable and Chern number is not present in magnetically-doped topological insulators is that not all interactions result in non-Hermitian self-energies with finite weight at $\omega=0$. For example, consider the effect of magnetic impurity scattering on the surface of a topological insulator. The anti-Hermitian component of the self-energy resulting from magnetic impurity scattering is of the form~\cite{philip_loss_2018}
\begin{equation}
	\Sigma = -i\Gamma_0|\omega|\text{sgn}(\text{Im }\omega),
\end{equation}
where $\Gamma_0$ quantifies the broadening induced by the magnetic impurity scattering. We immediately notice that the linear dependence of the self-energy on $|\omega|$ circumvents the discontinuity at $\omega=0$. The resulting Green function is continuous for all $\mathbf{k}$ and $\omega$ and is a legitimate homeomorphism from momentum space to $GL(N,\mathbb{C})$. Therefore, Eq.~\eqref{eq:IMdef} produces a quantized Hall conductivity, consistent with experimental results.

In summary, we have studied the connection between observables and topological invariants in non-Hermitian Chern insulators. We have analytically shown via field theoretical techniques that there exists a disconnect between the Chern number and the Hall conductivity in $(2+1)$-D non-Hermitian Hamiltonians, proving that Hermitian and non-Hermitian Chern insulators are not adiabatically connected to one another. We derived an exact formula for the Hall conductivity of generic two-level non-Hermitian systems that clearly demonstrates the disconnect from the Chern number. For the particular case of a non-Hermitian massive Dirac Hamiltonian, we showed that as broadening is introduced, the Hall conductivity deviates from its quantized value and the system develops a longitudinal conductivity. We have further shown that the disconnect between topology and observable may be extended to higher dimensional systems, specifically addressing $(4+1)$-D systems characterized by the second Chern number. Importantly, we have illustrated that our results are consistent with the experimental observations of the quantum anomalous Hall effect in magnetically-doped topological insulators. Our results demonstrate the necessity of reexamining perceived links between topology and the quantization of observables in non-Hermitian systems. 

\begin{acknowledgments}
M.R.H., T.M.P., and M.J.G. acknowledge financial support from the National Science Foundation (NSF) under CAREER Award ECCS-1351871 and from the Office of Naval Research (ONR) under Grant No. N00014-17-1-3012. M.R.H. and M.J.G. acknowledge financial support from the NSF under Grant No. DMR 17-10437. M.J.G. acknowledges financial support from the NSF under Grant No. DMR-1720633. M.R.H. acknowledges fruitful discussions with X.-L. Qi, B. Basa, and G. A. Hamilton.
\end{acknowledgments}

%

\end{document}


\title{Supplemental Material for ``Topology and Observables of the Non-Hermitian Chern Insulator''}

\author{Mark R. Hirsbrunner}\email{hrsbrnn2@illinois.edu}\affiliation{\UIUCPHYS}\affiliation{\UIUCMNTL}
\author{Timothy M. Philip}\altaffiliation[Current address: ]{\IBM}\affiliation{\UIUCMNTL}\affiliation{\UIUCECE}
\author{Matthew J. Gilbert}\affiliation{\UIUCMNTL}\affiliation{\UIUCECE}\affiliation{\STANFORDECE}

\maketitle

\section{Green functions of non-Hermitian Hamiltonians.}
For a Hermitian Hamiltonian $H_0(\mathbf{k})$, the single-particle retarded and advanced Green functions are~\cite{kadanoff_quantum_1962,bruus_many_2004}
\begin{align}
	G^{r/a}_0(\v k, E) 	&= \frac{1}{(E \pm i0^+)I - H_0(\v k)} \\
						&=  \sum_\alpha \frac{1}{E \pm i0^+ - \epsilon_\alpha(\v k) }
\end{align}
where $\epsilon_\alpha(\mathbf{k})$ is the energy of band $\alpha$ at momentum $\mathbf{k}$. The effects of interactions and openness (non-Hermiticity) can be accounted for in the single-particle Green functions via the inclusion of a self-energy $\Sigma^{r/a}(\mathbf{k},E)$, which modifies the retarded and advanced Green functions as
\begin{align}
	G^{r/a}(\v k, E) 	&= \left[(E \pm i0^+)I - H_0(\v k) - \Sigma^{r/a}(\v k, E)\right]^{-1} \\
						&=  \sum_\alpha \left[E \pm i0^+ - \epsilon_\alpha(\v k) - \text{Re}\,\Sigma^{r}_\alpha(\v k, E) \mp i \text{Im}\,\Sigma^{r}_\alpha(\v k, E)\right]^{-1}
\end{align}
where $\Sigma^{r/a}_\alpha = \left\langle\alpha\right|\Sigma^{r/a}\left|\alpha\right\rangle$ are the retarded and advanced self-energies. The final equality makes use of the relation $\Sigma^a=\Sigma^{r\dagger}$~\cite{bruus_many_2004,Anantram2008a,Pourfath2014}. Because the self-energy is constructed by integrating out interactions or subsystems, it is generally non-Hermitian~\cite{rotter_review_2015,rotter_non-hermitian_2009,bender_making_2007}. The spectral representation of the Green function makes clear the roles of the real and imaginary parts of the self-energy. The real part shifts the locations of the poles of the Green function along the real axis, while the imaginary part shifts the poles away from the real axis into the complex plane. The definition of the retarded and advanced Green functions requires that the poles be located in the lower and upper half-planes, respectively. Therefore, any proper self-energy must have the property $\text{Im }\Sigma_\alpha^r(\mathbf{k},E)<0$ for all $\mathbf{k}$ and $\alpha$.

Given an energy-independent self-energy, an effective non-Hermitian Hamiltonian can be constructed from the retarded Green function by grouping the Hamiltonian and retarded self-energy together~\cite{feshbach_unified_1958,roth_nonlocal_2009,rotter_review_2015}:
\begin{align}
	G^r(\v k, E)	&= \frac{1}{(E + i0^+)I - \left[H_0(\v k) + \Sigma^{r}(\v k)\right] } \\
					&= \frac{1}{(E + i0^+)I - H_\text{eff}(\v k)  }, \label{eq:Gr_effective}
\end{align}
\begin{equation}
	H_{\text{eff}}(\mathbf{k}) = H_0(\mathbf{k})+\Sigma^r(\mathbf{k}).
\end{equation}
Owing to the aforementioned restrictions on the imaginary part of the self-energy, the imaginary component of the energies of these non-Hermitian effective Hamiltonians must be negative.

In this work, we must construct Green functions from non-Hermitian Hamiltonians. A non-Hermitian Hamiltonian may be written as a sum of Hermitian and anti-Hermitian parts,
\begin{equation}
	H(\v{k}) = H_0(\v{k}) + \Gamma(\v{k}) ,\quad H_0=H_0^\dagger ,\quad \Gamma = -\Gamma^\dagger.
\end{equation}
By treating $\Gamma(\v{k})$ as a self-energy, the retarded and advanced Green functions can then be written as
\begin{equation}
	G^{r/a}(\mathbf{k},E) = \frac{1}{(E+i0^+)I - H_0(\mathbf{k}) \mp \Gamma(\mathbf{k})}.
\end{equation}

The zero-temperature limit of the Matsubara Green function can be defined in terms of the retarded and advanced Green functions as
\begin{equation}
	G(\mathbf{k},\omega) =
	\begin{cases}
		G^r(\mathbf{k},\omega) & \text{Im}\,\omega > 0 \\
		G^a(\mathbf{k},\omega) & \text{Im}\,\omega < 0,
	\end{cases}
\end{equation}
where we have changed variables from $E\in\mathbb{R}$ to $\omega\in\mathbb{C}$, as $G$ is defined for complex energies~\cite{karlsson_partial_2016}. The Green function may be written concisely as
\begin{equation}
	G(\mathbf{k},\omega) = \frac{1}{\omega - H_0(\mathbf{k}) - \Gamma(\mathbf{k})\text{sgn}(\text{Im }\omega)}.
\end{equation}

\section{Variation of the Hall conductivity}
A distortion in the Green function $G\rightarrow G+\delta G$ results in a variation
\begin{equation}
	G\partial_\mu G^{-1}\rightarrow -G(\partial_\mu G^{-1})\delta GG^{-1}-\partial_\mu (\delta G)G^{-1},
\end{equation}
where we have abbreviated $\partial/\partial p_\mu$ as $\partial_\mu$ and used the fact that $\delta G^{-1}=-G^{-1}\delta GG^{-1}$. Plugging this into the Hall conductivity gives the variation
\begin{equation}
	\begin{aligned}
		\delta \sigma_{xy}
		=&
		-\frac{e^2}{h}\frac{1}{24\pi^2}\epsilon^{\mu\nu\rho}\int\, d^3p\,\text{Tr}\left[(G\partial_\mu G^{-1}\delta GG^{-1}) (G\partial_\nu G^{-1}) (G\partial_\rho G^{-1})\right] \\
		&
		-\frac{e^2}{h}\frac{1}{24\pi^2}\epsilon^{\mu\nu\rho}\int\, d^3p\,\text{Tr}\left[(\partial_\mu \delta GG^{-1}) (G\partial_\nu G^{-1}) (G\partial_\rho G^{-1})\right].
	\end{aligned}
\end{equation}
Using the cyclic properties of the trace and inserting an identity $GG^{-1}=1$ yields
\begin{equation}
	\begin{aligned}
		\delta \sigma_{xy}
		=&
		-\frac{e^2}{h}\frac{1}{24\pi^2}\epsilon^{\mu\nu\rho}\int\, d^3p\,\text{Tr}\left[(\partial_\mu G^{-1}\delta G)(\partial_\nu G^{-1} G)(\partial_\rho G^{-1}G)\right] \\
		&
		-\frac{e^2}{h}\frac{1}{24\pi^2}\epsilon^{\mu\nu\rho}\int\, d^3p\,\text{Tr}\left[(G^{-1}\partial_\mu \delta G)(\partial_\nu G^{-1}G)(\partial_\rho G^{-1}G)\right].
	\end{aligned}
\end{equation}
The first term in the two traces may be written as a single derivative,
\begin{equation}
	\delta \sigma_{xy}=-\frac{e^2}{h}\frac{1}{24\pi^2}\epsilon^{\mu\nu\rho}\int\, d^3p\,\text{Tr}\left[\partial_\mu (G^{-1}\delta G)(\partial_\nu G^{-1} G)(\partial_\rho G^{-1}G)\right].
\end{equation}
We then integrate by parts:
\begin{equation}
	\begin{aligned}
		\delta \sigma_{xy}
		=&
		-\frac{e^2}{h}\frac{1}{24\pi^2}\epsilon^{\mu\nu\rho}\int\, d^3p\,\partial_\mu \text{Tr}\left[(\delta G)(\partial_\nu G^{-1} G)(\partial_\rho G^{-1})\right] \\
		&
		+\frac{e^2}{h}\frac{1}{24\pi^2}\epsilon^{\mu\nu\rho}\int\, d^3p\,\text{Tr}\left[(G^{-1}\delta G)\partial_\mu (\partial_\nu G^{-1} G)(\partial_\rho G^{-1}G)\right] \\
		&
		+\frac{e^2}{h}\frac{1}{24\pi^2}\epsilon^{\mu\nu\rho}\int\, d^3p\,\text{Tr}\left[(G^{-1}\delta G)(\partial_\nu G^{-1} G)\partial_\mu (\partial_\rho G^{-1}G)\right].
	\end{aligned}
\end{equation}
Integration by parts has introduced second derivative terms, $\partial_\mu(\partial_\nu G^{-1}G)$. By the chain rule, the newly introduced second derivative terms equal
\begin{equation}
	\partial_\mu(\partial_\nu G^{-1}G) 	= (\partial_\mu\partial_\nu G^{-1})G + \partial_\nu G^{-1}\partial_\mu G. \label{eq:chain_rule}
\end{equation}
Contracting the first term in eqn.~\ref{eq:chain_rule} with the Levi-Civita symbol gives zero, as it is symmetric in $\mu$ and $\nu$, leaving only the second term. The variation of $\sigma_{xy}$ is then
\begin{equation}
	\begin{aligned}
		\delta \sigma_{xy}
		=&
		-\frac{e^2}{h}\frac{1}{24\pi^2}\epsilon^{\mu\nu\rho}\int\, d^3p\,\partial_\mu \text{Tr}\left[(\delta G)(\partial_\nu G^{-1} G)(\partial_\rho G^{-1})\right] \\
		&
		+\frac{e^2}{h}\frac{1}{24\pi^2}\epsilon^{\mu\nu\rho}\int\, d^3p\,\text{Tr}\left[(\delta G)(\partial_\nu G^{-1}\partial_\mu G)(\partial_\rho G^{-1})\right] \\
		&
		+\frac{e^2}{h}\frac{1}{24\pi^2}\epsilon^{\mu\nu\rho}\int\, d^3p\,\text{Tr}\left[(\delta G)(\partial_\nu G^{-1} G)(\partial_\rho G^{-1}\partial_\mu GG^{-1})\right].
	\end{aligned}
\end{equation}
Using the identity $\partial_\mu G^{-1}=-G^{-1}\partial_\mu G G^{-1}$ allows us to rewrite this as
\begin{equation}
	\begin{aligned}
		\delta \sigma_{xy}
		=&
		-\frac{e^2}{h}\frac{1}{24\pi^2}\epsilon^{\mu\nu\rho}\int\, d^3p\,\partial_\mu \text{Tr}\left[(\delta G)(\partial_\nu G^{-1} G)(\partial_\rho G^{-1})\right] \\
		&
		+\frac{e^2}{h}\frac{1}{24\pi^2}\epsilon^{\mu\nu\rho}\int\, d^3p\,\text{Tr}\left[(\delta G)(\partial_\nu G^{-1}\partial_\mu G)(\partial_\rho G^{-1})\right] \\
		&
		+\frac{e^2}{h}\frac{1}{24\pi^2}\epsilon^{\mu\nu\rho}\int\, d^3p\,\text{Tr}\left[(\delta G)(\partial_\nu G^{-1}\partial_\rho G)(\partial_\mu G^{-1})\right].
	\end{aligned}
\end{equation}
The last two terms are identical up to an exchange of $\mu$ and $\rho$ and thus cancel when contracted with the Levi-Civita symbol. This leaves only the total derivative
\begin{equation}
	\delta \sigma_{xy} = -\frac{e^2}{h}\frac{1}{24\pi^2}\epsilon^{\mu\nu\rho}\int\, d^3p\,\partial_\mu \text{Tr}\left[(\delta G)(\partial_\nu G^{-1} G)(\partial_\rho G^{-1})\right].
\end{equation}
Explicitly writing out the frequency derivatives gives
\begin{equation}
	\begin{aligned}
		\delta \sigma_{xy}
		=&
		-\frac{e^2}{h}\frac{\epsilon^{ij}}{24\pi^2}\int\, d^3p\,\partial_0 \text{Tr}\left[\delta G(\partial_i G^{-1} G)(\partial_j G^{-1})\right] \\
		&
		+\frac{e^2}{h}\frac{\epsilon^{ij}}{24\pi^2}\int\, d^3p\,\partial_i \text{Tr}\left[\delta G(\partial_0 G^{-1} G)(\partial_j G^{-1})\right] \\
		&
		-\frac{e^2}{h}\frac{\epsilon^{ij}}{24\pi^2}\int\, d^3p\,\partial_i \text{Tr}\left[\delta G(\partial_j G^{-1} G)(\partial_0 G^{-1})\right].
	\end{aligned}
\end{equation}
Consider the representative non-Hermitian Green function
\begin{align}
	G(\mathbf{k},\omega) 	&= \left[(\omega + i\Gamma_0(\omega,\mathbf{k})\,\text{sgn}(\text{Im}\,\omega))I - H_0(\mathbf{k})\right]^{-1}.
\end{align}
Here the self-energy is an anti-Hermitian diagonal matrix with some dependence on $\mathbf{k}$ and $\omega$, and $\Gamma_0$ always a positive real number. The divergence theorem can be applied to the last two lines of the variation yielding zero, but the possibility of a discontinuity in the Green function prevents the divergence theorem from applying to the first line. Using the substitution $\omega' = \omega + i\Gamma_0(\omega,\mathbf{k})\,\text{sgn}(\text{Im}\,\omega)$, the Green function becomes
\begin{equation}
	G(\mathbf{k},\omega') = \left[(\omega' + \text{sgn}(\text{Im}\,\omega) i0^+)I - H_0(\mathbf{k})\right]^{-1},
\end{equation}
which has the same form as the bare Green function $G_0(\mathbf{k},\omega)$ with no self-energy. Inserting this into $\delta \sigma_{xy}$ gives
\begin{align}
	\delta \sigma_{xy}
	=&
	-\frac{e^2}{h}\frac{\epsilon^{ij}}{24\pi^2}\left(\int_{-i\infty}^{-i\Gamma_0(0,\mathbf{k})}\,d\,\omega' + \int^{i\infty}_{i\Gamma_0(0,\mathbf{k})}\,d\,\omega'\right)\partial_{\omega'}\int d^2k\, \text{Tr}\left[\delta G(\partial_i G_0^{-1} G_0)(\partial_j G_0^{-1})\right] \\
	=&
	\frac{e^2}{h}\frac{\epsilon^{ij}}{24\pi^2}\left.\left\{\int d^2k\, \text{Tr}\left[\delta G(\partial_i G_0^{-1} G_0)(\partial_j G_0^{-1})\right]\right\}\right|_{\omega'=-i\Gamma_0(0,\mathbf{k})}^{\omega'=i\Gamma_0(0,\mathbf{k})},
\end{align}
which is only identically zero when $\Gamma_0(\mathbf{k},0)=0$. Non-Hermitian Green functions always have $\Gamma_0(\mathbf{k},0)\neq0$, thus $\sigma_xy$ cannot be a topological invariant in non-Hermitian systems.

\section{Generic non-Hermitian two-level system Hall conductivity}
We now derive the Hall conductivity of a generic non-Hermitian Hamiltonian
\begin{equation}
		H = \left(d_0(\v{k}) + i\Gamma_0(\v{k})\right)\sigma_0+\left(\v{d}(\v{k}) + i\v{\Gamma}(\v{k})\right)\cdot\v{\sigma}.
\end{equation}
The Green function is
\begin{equation}
	G = \left[\left(\omega-d_0(\v{k})-i\text{sgn}(\text{Im}\,\omega)\Gamma_0(\v{k})\right)\sigma_0 - \left(\v{d}(\v{k}) + i\text{sgn}(\text{Im}\,\omega)\v{\Gamma}(\v{k})\right)\cdot\v{\sigma}\right]^{-1}.
\end{equation}
Plugging this Green function into the Hall conductivity formula (Eq.~(4) in the main text) and performing the substitution $\v{D}=\left(\v{d}+i\text{sgn}(\text{Im }\omega)\v{\Gamma}\right)$ yields
\begin{equation}
	\sigma_{xy} = \frac{e^2}{h}\frac{i}{2\pi^2}\int_{-i\infty}^{i\infty}d\omega\int d^2k\frac{\v{D}\cdot\left\{\frac{\partial \v{D}}{\partial k_x}\times\frac{\partial \v{D}}{\partial k_y}\right\}}{\left[(\omega - d_0(\v{k}) - i\text{sgn}(\text{Im}\,\omega)\Gamma_0(\v{k}))^2-\v{D}^2\right]^2}.
\end{equation}
From this point on we suppress any momentum dependence. We simplify this expression by splitting the frequency integral into the positive and negative halves of the imaginary axis to eliminate the $\text{sgn}(\text{Im }\omega)$, and utilize the substitutions $b_0=d_0+i\Gamma_0$ and $\v{b}=\left(\v{d}+i\v{\Gamma}\right)$:
\begin{align}
	\sigma_{xy} =& \frac{e^2}{h}\frac{i}{2\pi^2}\int d^2k\left[\int_{-i\infty}^{0}d\omega\frac{\v{b}^*\cdot\left(\frac{\partial \v{b}^*}{\partial k_x}\times\frac{\partial \v{b}^*}{\partial k_y}\right)}{\left((\omega - b_0^*)^2-\v{b}^*\cdot\v{b}^*\right)^2} + \int^{i\infty}_{0}d\omega\frac{\v{b}\cdot\left(\frac{\partial \v{b}}{\partial k_x}\times\frac{\partial \v{b}}{\partial k_y}\right)}{\left((\omega - b_0)^2-\v{b}\cdot\v{b}\right)^2}\right].
\end{align}
Explicitly performing the frequency integration and carefully taking the limits at $\pm i\infty$ yields~\footnote{We have chosen the branch cuts of the arctan function to be $(i,i\infty)$ and $(-i,-i\infty)$.}
\begin{align}
\sigma_{xy} =	& -\frac{e^2}{h}\int \frac{d^2k}{2\pi^2}\text{Re}\left[\hat{b}\cdot\left(\frac{\partial \hat{b}}{\partial k_x}\times\frac{\partial \hat{b}}{\partial k_y}\right)\left(\frac{\pi}{2}\text{sgn}(\text{Re } b) - \frac{ibb_0}{b^2-b_0^2} - i\text{ arctanh}\left(\frac{b_0}{b}\right)\right)\right],
\end{align}
where $b=\sqrt{\v{b}\cdot\v{b}}$ and $\hat{b}=\v{b}/b$. This is Eq.~(13) in the main text. We note that the choice of positive or negative square root does not affect the calculation. 

\section{Non-Hermitian Massive Dirac Hall Conductivity}
Here we explicitly calculate the Hall conductivity of the non-Hermitian massive Dirac Hamiltonian in $(2+1)$-D using Eq.~(4) from the main text. The Hamiltonian is defined as 
\begin{equation}
H(\v{k}) = -(\mu + i\Gamma_0)\sigma_0 + \nu_F\v{k}\cdot\v{\sigma} + M\sigma_z.
\end{equation}
Using the Hamiltonian, we may assemble the Matsubara Green function,
\begin{equation}
	\begin{aligned}
		G 	&= \left[\left(\mu + \omega + i\Gamma_0\text{sgn}(\text{Im}\,\omega)\right)\sigma_0 - \nu_F\left(k_x\sigma_x + k_y\sigma_y\right) - M\sigma_z\right]^{-1}.
	\end{aligned}
\end{equation}
The self-energy associated with the non-Hermiticity is absorbed into $\omega$ by the substitution $\omega'=\omega + i\Gamma_0\text{sgn}(\text{Im}\,\omega)$, giving a simplified expression for the Green function
\begin{equation}
	G(\omega')=\frac{\left(\mu+\omega'\right)\sigma_0+\nu_F\left(k_x\sigma_x + k_y\sigma_y\right)+M\sigma_z}{\left(\omega'+\mu\right)^2 - M^2 - \nu_F^2k^2}.
\end{equation}
In order to calculate the Hall conductivity, we require the derivatives of $G^{-1}$:
\begin{equation}
	\frac{\partial G^{-1}}{\partial \omega'} = \sigma_0,\,
	\frac{\partial G^{-1}}{\partial k_x} = \nu_F\sigma_x,\,
	\frac{\partial G^{-1}}{\partial k_y} = \nu_F\sigma_y.
\end{equation}
Having assembled the constituent components, we may evaluate the trace and sum over the indices, resulting in
\begin{equation}
\frac{e^2}{h}\frac{\epsilon^{\mu\nu\rho}}{24\pi^2}\text{Tr}\left[G\frac{\partial G^{-1}}{\partial p_\mu} G\frac{\partial G^{-1}}{\partial p_\nu} G\frac{\partial G^{-1}}{\partial p_\rho}\right] = \frac{e^2}{h}\frac{iM\nu_F^2}{2\pi^2\left(\left(\omega' + \mu\right)^2 - M^2 -  \nu_F^2k^2\right)^{2}}.
\end{equation}
The substitution $\omega\rightarrow\omega'$ changes the $\omega$ integration from the entire imaginary axis, $\omega=(-i\infty,i\infty)$, to the subset $\omega'=(-i\infty,-i\Gamma_0)\cup(i\Gamma_0,i\infty)$. Explicitly evaluating the expression for the Hall conductivity then yields
\begin{equation}
	\begin{aligned}
		\sigma_{xy} 	&= \frac{e^2}{h}\frac{iM\nu_F^2}{2\pi^2}\left(\int_{-i\infty}^{-i\Gamma_0} d\omega' + \int^{i\infty}_{i\Gamma_0} d\omega'\right)\int_0^\infty dk\,2\pi k\left(\left(\omega' + \mu\right)^2 - M^2 - \nu_F^2k^2\right)^{-2} \\
	 					&= -\frac{e^2}{h}\left(\frac{1}{2} - \frac{1}{2\pi}\left[\arctan\left(\frac{\Gamma_0+i\mu}{M}\right) + \arctan\left(\frac{\Gamma_0-i\mu}{M}\right)\right]\right) \\
	 					&= -\frac{e^2}{h}\frac{M}{2\pi|M|}\left[\frac{\pi}{2} - \arctan\left(\frac{\mu^2+\Gamma_0^2-M^2}{2\Gamma_0|M|}\right)\right].
	\end{aligned}
\end{equation}

%